\documentclass[lettersize,journal]{IEEEtran}

\usepackage{amsmath}
\usepackage{amsfonts}
\usepackage{algorithmic}
\usepackage{graphicx}
\usepackage{subfigure}
\usepackage{textcomp}
\usepackage{xcolor}
\usepackage{booktabs}
\usepackage{colortbl}
\usepackage{threeparttable}
\usepackage{color}
\usepackage{multirow}
\usepackage{url}
\usepackage{enumitem}
\newcommand{\argmin}{\arg\!\min}

\begin{document}

\title{Breaking the Curse of Knowledge: \\ Towards Effective Multimodal Recommendation using Knowledge Soft Integration} 

\author{Kai Ouyang*, Chen Tang*, Zenghao Chai, Wenhao Zheng, Xiangjin Xie, \\ Xuanji Xiao, Zhi Wang$^{\dagger}$,~\IEEEmembership{Senior Member,~IEEE}
\IEEEcompsocitemizethanks{
\IEEEcompsocthanksitem Kai Ouyang, Xiangjin Xie, and Zhi Wang are with Tsinghua University, China. 
Chen Tang is with MMLab, The Chinese University of Hong Kong, Hong Kong. 
Zenghao Chai is with National University of Singapore, Singapore. 
Wenhao Zheng is with Microsoft AI Asia, China. 
\\
Email: plucky.ouykai@gmail.com, chentang@link.cuhk.edu.hk, and wangzhi@sz.tsinghua.edu.cn. 

\IEEEcompsocthanksitem $^{*}$Equal contributions. ~~\textsuperscript{\dag}Corresponding author.
}
}

\maketitle

\begin{abstract}
A critical challenge in contemporary recommendation systems lies in effectively leveraging multimodal content to enhance recommendation personalization. Although various solutions have been proposed, most fail to account for discrepancies between knowledge extracted through isolated feature extraction and its application in recommendation tasks. Specifically, multimodal feature extraction does not incorporate task-specific prior knowledge, while downstream recommendation tasks typically use these features as auxiliary information. This misalignment often introduces biases in model fitting and degrades performance, a phenomenon we refer to as the curse of knowledge.
To address this challenge, we propose a knowledge soft integration framework designed to balance the utilization of multimodal features with the biases they may introduce. The framework, named \textbf{K}nowledge \textbf{S}oft \textbf{I}ntegration (KSI), comprises two key components: the Structure Efficient Injection (SEI) module and the Semantic Soft Integration (SSI) module.
The SEI module employs a Refined Graph Neural Network (RGNN) to model inter-modal correlations among items while introducing a regularization term to minimize redundancy in user and item representations. In parallel, the SSI module utilizes a self-supervised retrieval task to implicitly integrate multimodal semantic knowledge, thereby enhancing the semantic distinctiveness of item representations.
We conduct comprehensive experiments on three benchmark datasets, demonstrating KSI's effectiveness. 
Furthermore, these results underscore the ability of the SEI and SSI modules to reduce representation redundancy and mitigate the curse of knowledge in multimodal recommendation systems.
\end{abstract}

\section{Introduction}
\label{sec:intro}

\IEEEPARstart{A}{s} multimodal content increasingly permeates online platforms, multimodal recommendation has gained significant attention, leveraging diverse content formats to identify items aligned with user interests. This approach has demonstrated particular success across e-commerce, instant video, and social media applications.

Effective feature extraction is critical to enhance multimodal recommendation performance. To increase feature accuracy while minimizing computational demands, most current multimodal recommendation methods~\cite{VBPR_he2016vbpr,LOGO_cao2020hashtag,MMGCN_wei2019mmgcn,MKGAT_sun2020multi,GRCN_yinwei2021grcn,LATTICE_zhang2021mining} treat feature extraction as an independent preprocessing stage, often using general pre-trained models to generate semantic embeddings from visual and textual modalities without integrating knowledge specific to recommendation tasks. 
For example, methods like MMGCN~\cite{MMGCN_wei2019mmgcn} and RGCN~\cite{GRCN_yinwei2021grcn} employ pre-trained models such as ResNet50 for visual feature extraction. In subsequent recommendation tasks, prior studies typically apply simple fusion techniques (e.g., concatenation or attention mechanisms~\cite{ACF_chen2017attentive}) to incorporate these features into the model. Specifically, MMGCN~\cite{MMGCN_wei2019mmgcn} aggregates features from items a user has interacted with, directly embedding them into user representations.

Multimodal recommendations generally outperform traditional recommendation methods (e.g., NGCF~\cite{NGCF_wang2019neural}, LightGCN~\cite{LightGCN_he2020lightgcn}) by leveraging diverse data sources. However, using multimodal features extracted independently of recommendation task relevance can introduce what we define as the \textit{curse of knowledge}~\cite{cok_camerer1989curse}. 
This phenomenon fundamentally stems from the cognitive bias and attention decay of the model after introducing massive multi-modal information, which manifests in two critical ways:
\textbf{(a) Cognitive Bias from Uncorroborated Modalities:} When modalities are used directly without cross-validation, single-modal representations may encode misleading signals. For instance, in \textbf{Example 1}~\label{example:y}, two items sharing identical iPhone images (a phone and its case) would be indistinguishable if processed through isolated visual feature extraction. However, their textual descriptions (``64GB iPhone 15'' vs. ``silicone iPhone case'') contain discriminative attributes. Direct embedding propagates this bias into representations, compromising model integrity.
\textbf{(b) Attention Decay from Irrelevant Features:} Raw multimodal data inevitably contains task-irrelevant dimensions that divert model focus. In \textbf{Example 2}~\label{example:x}, a user comparing iPhone prices cares little about artistic product images, yet conventional fusion methods treat all visual/textual features equally. This dilutes critical attributes (price/chip) while amplifying noise from decorative details (e.g., background patterns), degrading recommendation specificity.
Collectively, these limitations (direct embedding of uncorroborated features and over-reliance on irrelevant dimensions) constitute the \textit{curse of knowledge}, ultimately impairing recommendation performance.

To address the above issues, we design the KSI framework, which resolves these cognitive biases through \textit{indirect utilization and refinement of knowledge}: 
For Cognitive Bias from Uncorroborated Modalities, we propose a knowledge soft-injection strategy, incorporating task-relevant structure and semantic information from multimodal content into recommendation models. 
For Cognitive Bias from Uncorroborated Modalities, we enhance representation space utilization efficiency, fortifying the recommendation model’s robustness in handling diverse multimodal inputs. 
Our proposed framework, \textbf{K}nowledge \textbf{S}oft \textbf{I}ntegration (KSI), achieves this through two principal components: the Structure Efficient Injection (SEI) module and the Semantic Soft Integration (SSI) module. Together, these modules facilitate the soft integration of multimodal knowledge. 

KSI circumvents these by \textit{indirect utilization and refinement of knowledge}—extracting stable structural relationships through RGNN's redundancy suppression and grounding them via contrastive semantic alignment. This ensures task-aware knowledge integration without inheriting raw feature biases.
The SEI module, through a Refined Graph Neural Network (RGNN), captures structural information from multimodal inputs, focusing on inter-item similarities and distinctions that remain relatively stable through feature extraction. RGNN also reduces redundancy in information aggregation, enhancing both transmission efficiency and the expressiveness of item and user representations. 
The SSI module introduces a self-supervised contrastive learning task, which selectively captures semantic nuances in multimodal content by emphasizing relative rather than absolute similarity. This technique facilitates soft semantic integration, thereby addressing the curse of knowledge and optimizing downstream recommendation performance.
We conduct extensive experiments on various real-world datasets and demonstrate the effectiveness of our KSI framework compared with previous state-of-the-art (SOTA) methods. 
Moreover, we also conduct exhaustive ablations to verify the superiority of the proposed SEI module, SSI module, and RGNN.

In summary, the core contributions of this work are:
\begin{itemize} 
\item We highlight the \textit{curse of knowledge} phenomenon in multimodal recommendation and propose KSI to balance multimodal feature utilization with mitigation of this issue. 
\item We introduce the Refined Graph Neural Network (RGNN) to efficiently extract structural information from multimodal features, enhancing representation expressiveness. 
\item We develop the self-supervised Semantic Soft Integration (SSI) module, enabling more robust use of semantic information from multimodal content. 
\item KSI achieves consistent improvements over all baselines, with gains of 2.07\%-5.46\%. Further ablation experiments confirm the effectiveness of the SEI and SSI modules. 
\end{itemize}

\begin{figure*}[t]
\centering
\includegraphics[width=\textwidth]{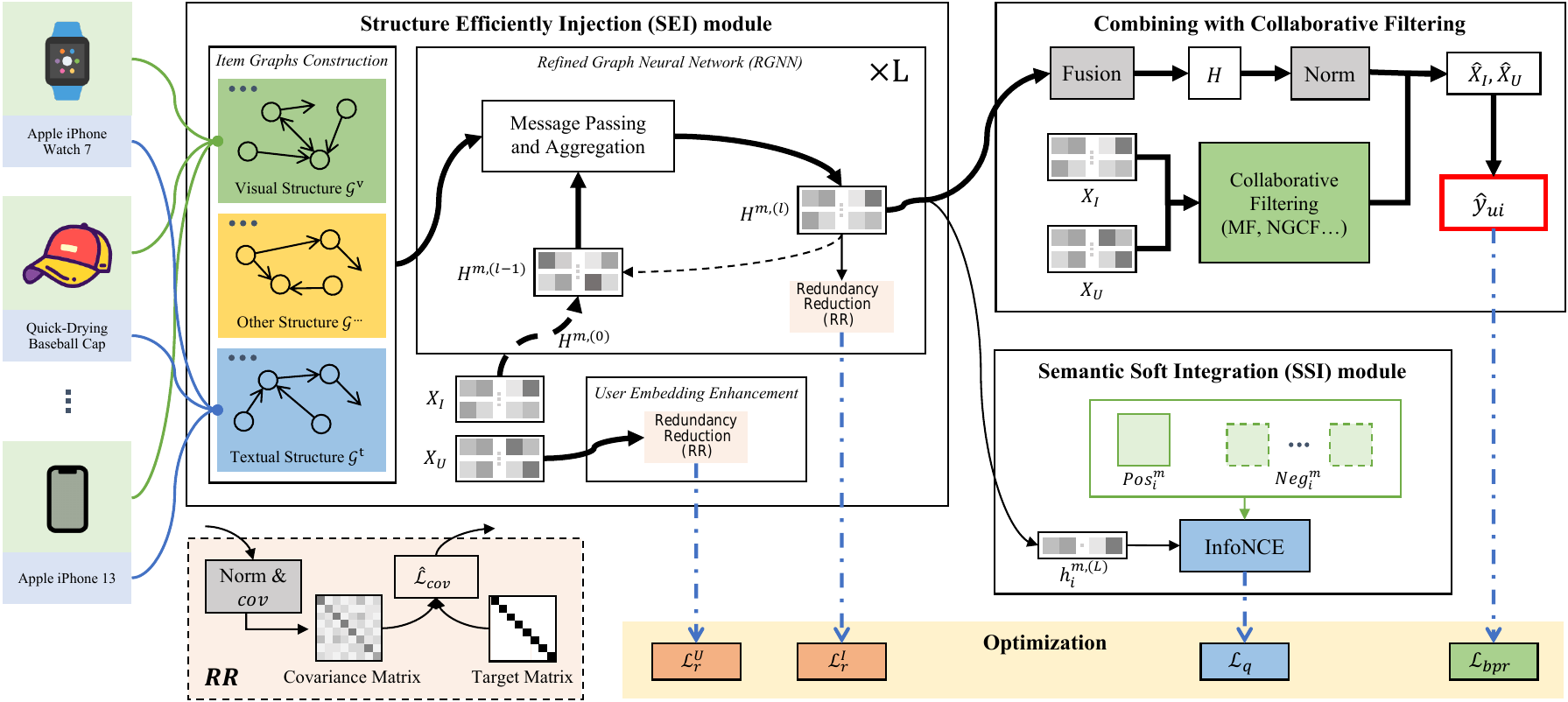} 
\caption{
Illustration of our proposed framework, Knowledge Soft
Integration (KSI). Bold paths are used to denote the backbone network. 
}
\label{fig:architecture}
\end{figure*}

\section{Related work}
\label{sec:rw}
In this section, we retrospect the existing works related to our research, including multimodal recommendation, graph neural networks, and self-supervised learning.

\subsection{Multimodal Recommendation}
In real-world recommendation scenarios, sparse user-item interactions often limit the performance of recommendation systems that rely on collaborative filtering (CF)~\cite{ICDE_khawar2019modeling}. Consequently, the multimodal recommendation has garnered substantial interest as a means to enhance these systems\cite{iccv_veit2015learning}.

\subsubsection{Traditional Methods}
Early multimodal recommendation frameworks~\cite{deepstyle_liu2017deepstyle,vecf_chen2019personalized,dvbr_kang2017visually} extended collaborative filtering (CF) by integrating multimodal content as auxiliary information in addition to basic user-item interactions. For instance, VBPR~\cite{VBPR_he2016vbpr} augmented matrix factorization by incorporating visual features to enhance recommendation accuracy. Similarly, DVBPR~\cite{dvbr_kang2017visually} jointly trained image representations and recommendation model parameters, enabling better utilization of visual content. DeepStyle~\cite{deepstyle_liu2017deepstyle} modeled item style features and user preferences by extracting category information from visual representations. ACF~\cite{ACF_chen2017attentive} introduced an attention mechanism to infer user preferences from implicit feedback (e.g., photo likes, video views, and song downloads) combined with multimodal content. Furthermore, VECF~\cite{vecf_chen2019personalized} improved recommendation specificity by modeling users' diverse interests in various images and comments.

\subsubsection{Graph-based Methods}
The recent adoption of Graph Neural Networks (GNNs) in recommendation systems~\cite{gnn0_kipf2016semi,gnn1_velivckovic2017graph,gnn2_hu2019hierarchical,wang2019neural,LightGCN_he2020lightgcn,graph1_wu2019session,graph2_yu2020tagnn} has also extended to multimodal recommendations~\cite{MMGCN_wei2019mmgcn,GRCN_yinwei2021grcn,HUIGN_wei2021hierarchical}. GNNs effectively model user-item interactions as graphs while integrating multimodal features to enhance item embeddings. MMGCN~\cite{MMGCN_wei2019mmgcn} introduced a graph-based approach to represent multimodal interactions, pioneering this field. GRCN~\cite{GRCN_yinwei2021grcn} mitigated false-positive feedback by pruning noisy edges in interaction graphs, while MMGCL~\cite{MMGCL_yi2022multi} incorporated graph contrastive learning through modality-specific edge dropout and masking. SLMRec~\cite{SLMRec_tao2022self} leveraged multimodal data augmentation techniques, such as multi-modal pattern augmentation, to enrich content representations.
Additionally, HUIGN~\cite{HUIGN_wei2021hierarchical} proposed a hierarchical intent modeling approach, achieving significant gains in recommendation performance. LATTICE~\cite{LATTICE_zhang2021mining} explored latent structures extracted from various modalities and transferred these to downstream recommendation models. Unlike methods that embed multimodal representations directly, LATTICE alleviated issues related to the curse of knowledge by focusing on latent structural information.
\textbf{BM3~\cite{BM3_zhou2023bootstrap}} eliminates negative sampling in BPR loss through bootstrapped contrastive views, reducing computational overhead while maintaining performance. \textbf{DiffMM~\cite{DiffMM_jiang2024diffmm}} integrates graph diffusion models with cross-modal contrastive learning, dynamically aligning multimodal features with collaborative signals. Our results demonstrate KSI's superiority across all categories, underscoring its ability to balance structural and semantic knowledge.

\subsection{Graph Neural Network (GNN)}
GNNs have shown strong capabilities in processing data with graph-structured, making them widely used in various domains, such as link prediction~\cite{lk0_zhang2018link,lk1_zhang2020learning}, information retrieval~\cite{ir0_zhang2019neural,ir1_xiong2017explicit}, among others. Recently, substantial research~\cite{HGNN_feng2019hypergraph,HetGNN_zhang2019heterogeneous,MAGNN_fu2020magnn,DGNN_manessi2020dynamic} has focused on optimizing GNNs to achieve more efficient and reliable message passing and aggregation. 

For instance, GraphSAGE~\cite{graphsage_hamilton2017inductive} generates node embeddings by sampling and aggregating features from local neighborhoods, allowing for scalable inductive learning. The GCN model~\cite{GCN_kipf2016semi} proposes a method to learn hidden layer representations that capture both the local graph structure and node features, while GAT~\cite{GAT_velivckovic2017graph} incorporates attention mechanisms into propagation, enabling nodes to selectively attend to their neighbors during embedding computation. LightGCN~\cite{LightGCN_he2020lightgcn} further simplifies this approach by removing both linear transformations and non-linear activation functions, enhancing computational efficiency without sacrificing performance.

However, because GNNs iteratively aggregate information from each node’s neighborhood to compute embeddings, this mechanism exhibits a cascading effect: small graph noise propagates through neighborhoods, impacting multiple node embeddings. Furthermore, aggregating neighboring information leads to inevitable information fusion and collision, introducing noise and representation redundancy.

To address these issues, we propose the Refined Graph Neural Network (RGNN). This approach is designed to enhance aggregation accuracy and minimize noise propagation by specifically refining information aggregation during message passing. This improves the robustness and expressiveness of the generated node representations.

\subsection{Self-Supervised Learning (SSL)}
SSL is an advancing technique that derives representations from self-generated supervision signals within raw data, eliminating the need for annotated labels. Among various SSL methods, contrastive learning is promising to achieve robust and discriminative representations by drawing positive sample pairs closer in representation space while distancing negative pairs. 
SSL has achieved notable success across diverse research fields, ranging from computer vision (CV)~\cite{SimCLR_chen2020simple,MoCo_he2020momentum} to natural language processing (NLP)~\cite{nlp0_wu2020clear,nlp1_gao2021simcse}.

In the realm of recommendation systems, SSL has shown strong potential. For instance, SGL~\cite{SSL0_wu2021self} integrates self-supervised auxiliary tasks within the CF frameworks, enhancing both the accuracy of GNNs and improving their robustness. S$^3$-Rec~\cite{SSL2_zhou2020s3} further applies contrastive learning in sequential recommendation (SR) tasks. CLCRec~\cite{SSL3_wei2021contrastive} leverages mutual information maximization between collaborative signals and contents to address the cold-start problem.

\section{Method}
\label{sec:me}
In this section, we first show the problem definition and its corresponding mathematical notation, and then introduce the Knowledge Soft Integration (KSI) framework in detail.

\subsection{Problem definition}
Define $\mathcal{U}(|\mathcal{U}|=N_{\mathcal{U}})$ as the set of users and $\mathcal{I}(|\mathcal{I}|=N_{\mathcal{I}})$ as the set of items. Each user $u \in \mathcal{U}$ is associated with a subset $\mathcal{I}^{u} \subseteq \mathcal{I}$ containing items with positive feedback, where each item $i \in \mathcal{I}^{u}$ has an assigned preference score $y_{ui} = 1$. Here, $N_{\mathcal{U}}$ and $N_{\mathcal{I}}$ indicate the total numbers of users and items, respectively. The input ID embeddings for user $u$ and item $i$ are represented as $x_{u}, x_{i} \in \mathbb{R}^{d}$, where $d$ corresponds to the embedding dimension. The complete embedding matrices for users and items are denoted as $X_{\mathcal{U}} \in \mathbb{R}^{N_{\mathcal{U}} \times d}$ and $X_{\mathcal{I}} \in \mathbb{R}^{N_{\mathcal{I}} \times d}$, respectively.

In addition to standard \textit{user-item} interactions, items incorporate multimodal features as content-based information. For any item $i$ in modality $m$, the modality-specific feature is expressed as $e^{m}{i} \in \mathbb{R}^{d_m}$, where $d_m$ is the dimensionality of the feature and $m \in \mathcal{M}$, with $\mathcal{M}$ representing the set of all modalities. The objective of multimodal recommendation is to effectively predict user preferences by ranking items for each user based on predicted preference scores $\hat{y}_{ui}$.

In this work, we specifically refer to the visual modality as $\text{v}$ and the textual modality as $\text{t}$.

\subsection{Architecture}
The Knowledge-Structured Integration (KSI) framework processes input data comprising \textit{user-item} historical interaction records and multimodal item features, producing as output the probability of a user's interaction with a specific item. The framework, illustrated in Figure~\ref{fig:architecture}, is organized into four primary components: 
\textbf{Structure Efficiently Injection (SEI) Module}: This module enhances structural information through a three-step process: (1) Item Graph Construction; (2) Refined Graph Neural Network; and (3) User Embedding Enhancement. 
\textbf{Semantic Soft Integration (SSI) Module}: This component leverages multimodal item content to extract semantic features, effectively improving item representations by capturing complementary information. 
\textbf{Integration with Collaborative Filtering (CF)}: User and item representations, both raw and enhanced, are incorporated into collaborative filtering algorithms to predict users' preferences for items. 
\textbf{Optimization}: The entire framework is fine-tuned using an objective function that balances a weighted combination of loss terms and regularization constraints, aiming to maximize predictive accuracy and model generalization.

\subsection{Structure Efficient Injection (SEI) Module} 
The integration of structural and semantic information is critical for effectively harnessing the knowledge embedded in multimodal data~\cite{LATTICE_zhang2021mining,SLMRec_tao2022self}. To address this, we introduce the SEI module, which incorporates a novel Refined Graph Neural Network (RGNN) for extracting and utilizing structural data from multimodal content. This approach facilitates efficient transfer of structural information by optimizing the representation space during node aggregation, while minimizing redundancy in the representations of users and items.

\subsubsection{Item Graphs Construction} \label{cig} To capture robust \textit{item-item} relationships, we initially construct a \textit{k}-NN modality-aware graph $\mathcal{G}^{m}$. The input to this layer consists of each item’s raw modality feature $e_{i}^{m}$, where $m \in \mathcal{M}$.

Following the collaborative filtering assumption~\cite{MF_rendle2012bpr}—that latent correlations between items are crucial for recommendation—we compute similarity between items based on multimodal content features. For simplicity and computational efficiency, we adopt cosine similarity~\cite{AMGCN_wang2020gcn} to quantify vector similarity as follows: \begin{equation} S_{ij}^{m} = \frac{(e_{i}^{m})^{\text{T}}e_{j}^{m}}{||e_{i}^{m}||\ ||e_{j}^{m}||}, \label{eq1} \end{equation} where $S_{ij}^{m}$ represents the scalar latent correlation between items $i$ and $j$, and $e^{m}$ denotes the raw modality features. The resulting similarity matrix $S^{m} \in \mathbb{R}^{N_{\mathcal{I}} \times N_{\mathcal{I}}}$ contains pairwise similarity scores, with negative values in $S^{m}$ set to zero.

To reduce computational complexity and mitigate noisy or irrelevant edges~\cite{noisy_chen2020iterative}, we apply sparsification to the adjacency matrix. Specifically, we perform \textit{k}-NN sparsification~\cite{KNN_chen2009fast} on $S^{m}$ by retaining only the edges with the top-\textit{k} similarity scores for each item $i$:
\begin{equation}
\begin{split}
    &\hat{S}_{ij}^{m}=
\left\{
\begin{array}{rcl}
S_{ij}^{m},      &      & S_{ij}^{m} \in \text{top-}k(S_{i,:}^{m}), \\
0,           &      & \text{otherwise},
\end{array} 
\right. 
\end{split}
\label{eq2}
\end{equation}
where $\hat{S}^{m} \in \mathbb{R}^{N_{\mathcal{I}} \times N_{\mathcal{I}}}$ represents the directed adjacency matrix, and $\text{top-}k(S_{i,:}^{m})$ selects the $k$ most similar items to item $i$.

To address issues such as exploding or vanishing gradients~\cite{Laplace_kipf2016semi}, we normalize the adjacency matrix as follows: \begin{equation} \mathcal{G}^{m} = (D^{m})^{-\frac{1}{2}} \hat{S}^{m} (D^{m})^{-\frac{1}{2}}, \label{eq3} \end{equation} where $D^{m} \in \mathbb{R}^{N_{\mathcal{I}} \times N_{\mathcal{I}}}$ is the diagonal degree matrix for $\hat{S}^{m}$, with $D_{ii} = \sum_{j}(\hat{S}_{ij}^{m})$. This produces the final modality-aware item graph $\mathcal{G}^{m} \in \mathbb{R}^{N{\mathcal{I}} \times N_{\mathcal{I}}}$, which represents the structural information for modality $m$. 

\subsubsection{Refined Graph Neural Network (RGNN)}
To utilize the extracted structural information and capture higher-order relationships, we design the Refined Graph Neural Network (RGNN) to enhance item representations by embedding \textit{item-item} affinities within the representation learning process. Specifically, RGNN enables each item to aggregate information from its first-order neighbors by propagating representations from connected items. This mechanism minimizes critical information loss during node aggregation, thereby supporting efficient embedding of structural information. By stacking multiple RGNN layers, the model effectively captures high-order relationships.
Basically, RGNN operates on the item graph $\mathcal{G}^{m}$ and the ID embedding matrix $X_{\mathcal{I}}$, generating an enhanced item representation matrix alongside a regularization term.

The primary objective of RGNN is to perform message passing and aggregation. Inspired by He \emph{et al.}~\cite{LightGCN_he2020lightgcn}, RGNN adopts a streamlined approach to message passing and aggregation, deliberately excluding feature transformations and non-linear activations to ensure computational efficiency.

\textbf{Passing and Aggregation Processes of Message.} The process within the $l$-th layer is formulated as follows: \begin{equation} h_{i}^{m,(l)}=\sum_{j\in \mathcal{N}_{i}} \mathcal{G}_{ij}^{m} h_{j}^{m,(l-1)}, \end{equation} where $h_{i}^{m,(l)} \in \mathbb{R}^{d}$ represents the modality $m$ embedding of item $i$ at the $l$-th layer, with $d$ as the dimensionality of the KSI embedding space, and $\mathcal{N}_{i}$ denotes the neighbor set of item $i$. The initial representations $h_{i}^{m,(0)}$ are set to the corresponding ID embeddings, $x_{i} \in X_{\mathcal{I}}$. For clarity, let $H^{m,(l)} = [h_{0}^{m,(l)}, \cdots, h_{N_{\mathcal{I}}}^{m,(l)}]$, representing the hidden item embeddings at layer $l$. 

\label{ref:rr} Furthermore, RGNN enhances the efficiency of information aggregation within message passing and node aggregation. Specifically, node aggregation requires effective information fusion, with higher numbers of neighbors implying an increased volume of information to be aggregated. However, the storage capacity of these representations is constrained by the dimensionality $d$. Given finite computational and storage resources, dimensions cannot be increased indefinitely; thus, each dimension must be fully optimized to maximize representational capacity.

\textbf{Redundancy Reduction (RR).}
A simple yet effective approach for improving the efficiency of the representation space is to minimize inter-dimensional correlations. In this study, we utilize the covariance matrix to measure the correlation across dimensions. Specifically, we begin by normalizing the representation matrix for each modality $m$ to enable precise computation: 
\begin{equation}
    \hat{H}^{m} = \frac{H^{m}}{\text{max}(||H^{m}||_{2})},
    \label{eq:pca_0}
\end{equation}
where $H^{m}$ is the input representation matrix.
For item representations, $H^{m} = H^{m,(l)} \in \mathbb{R}^{\mathcal{N}_{\mathcal{I}} \times d}$.
For user representations, $H^{m} = X_{\mathcal{U}} \in \mathbb{R}^{\mathcal{N}_{\mathcal{U}} \times d}$.
Then, we calculate the covariance matrix of $\hat{H}^{m}$:
\begin{equation}
    C^{m} = cov((\hat{H}^{m})^{T}). 
    \label{eq:pca_1}
\end{equation}
Here, $C^{m} \in \mathbb{R}^{d \times d}$ denotes the covariance matrix, with $cov$ representing the covariance computation operation. To minimize inter-dimensional correlation, we aim to make $C^{m}$ approximate the identity matrix as closely as possible:
\begin{equation}
    \hat{\mathcal{L}}_{cov}^{m} = \argmin_{C^{m}} \sum_{i=j} (C_{ij}^{m}-1)^{2} + \sum_{i \neq j} (C_{ij}^{m})^{2}.
    \label{eq:pca_2}
\end{equation}
We therefore define the Redundancy Reduction (RR) process according to Eq. \eqref{eq:pca_0}, Eq. \eqref{eq:pca_1}, and Eq. \eqref{eq:pca_2} as:
\begin{equation}
    \hat{\mathcal{L}}_{cov}^{m} = \text{RR}(H^{m}),
\end{equation}
where $\hat{\mathcal{L}}_{cov}^{m} \in \mathbb{R}$ denotes the final output regularization term. 

The core operation of RGNN at the $l$-th layer can be expressed as (with the Cross Product operator $\times$): 
\begin{equation}
\begin{split}
    H^{m,(l)} & = \mathcal{G}^{m} \times H^{m,(l-1)}, \\
    \hat{\mathcal{L}}_{cov}^{m,(l)} & = \text{RR}(H^{m,(l)}). \\
\end{split}
\end{equation}

The redundancy reduction process is also performed for $H^{m,(0)}$ (the matrix of input item ID embedding), \emph{i.e.},
$\hat{\mathcal{L}}_{cov}^{m,(0)} = \text{RR}(H^{m,(0)})$.
Therefore, by stacking $L$ layers, we obtain $L+1$ regularization terms $\{\hat{\mathcal{L}}_{cov}^{m,(0)},\cdots,\hat{\mathcal{L}}_{cov}^{m,(L)}\}$.
Then, we aggregate them:
\begin{equation}
    \mathcal{L}_{r}^{\mathcal{I}} = \frac{1}{L+1}  \sum_{m \in \mathcal{M}} \sum_{l=0}^{L} a_{m} \hat{\mathcal{L}}_{cov}^{m,(l)}\ . 
    \label{eq:loss0}
\end{equation}
Here, we introduce a learnable parameter $a_{m}$ for representing the modality-specific importance score, and $\mathcal{M} = \{\text{v}, \text{t}\}$ denotes the set of modalities. To control the regularization term’s scale, we use an averaging operation. The softmax function is applied to maintain normalization of the item representation $h_{i}$, ensuring that $\sum_{m \in \mathcal{M}} a_{m} = 1$.

The output of RGNN consists of $H^{m,(L)}$ and $\mathcal{L}_{r}^{\mathcal{I}}$, where $H^{m,(L)}$ captures the structural information of multimodal content, and $\mathcal{L}_{r}^{\mathcal{I}}$ reduces redundancy in item representations, enhancing the model's expressiveness.

\subsubsection{User Embedding Enhancement}
Given the critical role of the user ID embedding matrix in minimizing redundancy from \textit{user-item} interaction modeling, we then adopt the RR process to the user ID embedding matrix $X_{\mathcal{U}} \in \mathbb{R}^{\mathcal{N}_{\mathcal{U}} \times d}$ for reducing the redundancy of user representations: 
\begin{equation}
    \mathcal{L}_{r}^{\mathcal{U}} = \text{RR}(X_{\mathcal{U}}). 
    \label{eq:loss2} 
\end{equation}

The redundancy reduction (RR) process is theoretically grounded in the Maximum-Entropy Principle. By minimizing the covariance matrix's deviation from identity through our RR loss $\mathcal L_{\mathrm{RR}}$, we essentially maximize the differential entropy of the latent representations under an energy budget constraint. This mathematical equivalence ensures that each dimension becomes an uncorrelated, unit-variance factor - the optimal configuration for efficient information coding. This maximum-entropy perspective explains how RR simultaneously reduces redundancy while enhancing representation expressiveness.

\subsection{Semantic Soft Integration (SSI) Module}
\label{ref:ssi}
Structural information encapsulates the relationships between paired items, while semantic information derived from multimodal content provides unique insights into individual items. Overlooking either type of information would severely limit the model's capacity to generate expressive representations. To overcome this limitation, we propose the SSI module, a self-supervised mechanism explicitly designed to flexibly integrate the semantic features of multimodal content. The module takes as input the matrix $H^{m,(L)}$, and the raw modality feature matrix $E^{m}$, producing a loss function as its output to guide the integration process.

To address the dimensional disparity between raw modality features and the model's embedding size, we first align dimensions to enable efficient retrieval. We employ PCA~\cite{PCA_pearson1901liii}, a statistical technique for dimensionality reduction, to accomplish this. PCA effectively reduces dimensions while preserving spatial features and minimizing distortion~\cite{PCA0_he2010multimodal}: 
\begin{equation}
    \hat{E}^{m} = \text{PCA}(E^{m}, d), 
\end{equation} 
where $\hat{E}^{m} \in \mathbb{R}^{N_{\mathcal{I}} \times d}$ denotes the reduced-dimensionality modality feature matrix, where $d$ corresponds to the hidden embedding size of the model. Similarly, $E^{m} \in \mathbb{R}^{N_{\mathcal{I}} \times d_{m}}$ represents the original modality feature matrix, with $d_{m}$ indicating the embedding size of the raw modality features, and $m$ identifying the specific modality.

To balance computational cost and the effectiveness of the SSI module, we randomly sample $K$ images, text, or features from other modalities as negative examples for each item prior to each training epoch. 
The selection process can be formalized as:
\begin{equation}
    \text{Neg}_{i}^{m} = \text{Selection}_\texttt{random}(\hat{E}^{m}, K),
\end{equation}
where $\text{Neg}_{i}^{m}$ represents the negative sample set for item $i$, consisting of multimodal features that are not associated with item $i$. Notably, the final performance is not highly sensitive to the choice of $K$. Thus, in our implementation, we set $K$ empirically to $\frac{N_{\mathcal{I}}}{512}$.

We designate the modality features corresponding to item $i$ as the positive sample: 
\begin{equation}
    \text{Pos}_{i}^{m} = \{\hat{e}_{i}^{m}\},
\end{equation}
where $\text{Pos}_{i}^{m}$ is the positive sample set for item $i$, and $\hat{e}_{i}^{m} \in \hat{E}^{m}$ represents the modality features of item $i$. For simplicity, we define $P_{i}^{m} = \text{Neg}_{i}^{m} + \text{Pos}_{i}^{m}$ as the candidate pool for item $i$.

To capture the semantic correlations within multimodal content, we design a retrieval task aimed at aligning each item representation closely with its corresponding multimodal features while distancing it from unrelated ones. Using dot product as the similarity measure, we employ a contrastive loss function—specifically, InfoNCE~\cite{InfoNCE_oord2018representation,MoCo_he2020momentum}—to achieve this objective: 
\begin{equation}
    \mathcal{L}_{q} = - \sum_{m \in \mathcal{M}} \sum_{i=0}^{N_{\mathcal{I}}-1} a_{m} \log \frac{\exp(h_{i}^{m,(L)} \hat{e}_{i}^{m} / \tau)}{\sum_{\hat{e}^{m} \in P_{i}^{m}} \exp(h_{i}^{m,(L)} \hat{e}^{m} / \tau)},
    \label{eq:loss1}
\end{equation}
where $\tau$ is a temperature hyper-parameter~\cite{MoCo_he2020momentum}, $N_{\mathcal{I}}$ denotes the total number of items,  $h_{i}^{m,(L)} \in H^{m,(L)}$ is the output from the SEI module, and $a_{m}$ corresponds to the parameter defined in Eq.~\eqref{eq:loss0}. Additionally, since the negative examples for each item are refreshed at each epoch, the SSI module effectively aligns the item representation $h_{i}^{m,(L)}$ with its corresponding multimodal content across all modalities. 

Therefore, this enables the model to capture both semantic information, reflecting the affinity between corresponding multimodal features, and structural information, representing the relative similarity and dissimilarity among items within the multimodal content. By emphasizing relative similarity over absolute measures, the method alleviates the curse of knowledge issue. Fundamentally, the SSI module softly incorporates the semantic aspects of multimodal content into the model's training process.

\subsection{Combining with Collaborative Filtering}
Inspired by LATTICE~\cite{LATTICE_zhang2021mining}, KSI first derives item representations from multimodal features and then integrates them with downstream collaborative filtering (CF) models~\cite{MF_rendle2012bpr,NGCF_wang2019neural,LightGCN_he2020lightgcn} that capture \textit{user-item} interactions. This design makes KSI highly flexible, allowing it to function as a plug-and-play module compatible with any downstream CF approach. 

\textbf{Fusion.} Specifically, we first fuse the item representations $h_{i}^{m,(L)}$, which embed both structural and semantic information:
\begin{equation}
    h_{i} = \sum_{m \in \mathcal{M}} a_{m}h_{i}^{m,(L)},
\end{equation}
where $h_{i} \in \mathbb{R}^{d}$ denotes the fused representation, $h_{i}^{m,(L)} \in \mathbb{R}^{d}$ is the output from the SEI module, and $a_{m}$ corresponds to the parameters in Eq.~\eqref{eq:loss0} and Eq.~\eqref{eq:loss1}.

We denote the embeddings generated by collaborative filtering (CF) methods for users and items as $\hat{x}_{u}$ and $\hat{x}_{i} \in \mathbb{R}^{d}$, respectively. To enhance the item embedding $\hat{x}_{i}$, a normalized embedding $h{i}$ is added as follows:
\begin{equation}
\hat{x}_{i} = \hat{x}_{i} + \frac{h_{i}}{\text{max}(||h_{i}||_{2})}.
\end{equation}

The recommendation score is subsequently calculated as the inner product of the user embedding and the enhanced item embedding:
\begin{equation}
\hat{y}_{ui} = (\hat{x}_{u})^{T}\hat{x}_{i},
\end{equation}
where $\hat{y}_{ui}$ indicates the predicted probability of interaction between user $u$ and item $i$.

The optimization of the backbone network employs the Bayesian Personalized Ranking (BPR) loss~\cite{BPR_rendle2012bpr}, which encourages higher prediction scores for observed items compared to unobserved ones. The loss function is expressed as:
\begin{equation}
\mathcal{L}_{bpr} = \argmin \sum_{u \in \mathcal{U}} \sum_{i \in \mathcal{I}^{u}} \sum_{j \notin \mathcal{I}^{u}} - \ln \sigma (\hat{y}_{ui} - \hat{y}_{uj}),
\label{eq:loss3}
\end{equation}
where $\mathcal{I}^{u}$ refers to the set of positive items associated with user $u$, and $(u, i, j)$ represents triplets comprising a positive item $i \in \mathcal{I}^{u}$ and a negative item $j \notin \mathcal{I}^{u}$. The function $\sigma$ denotes the Sigmoid activation.

\subsection{Optimization}
At this stage, we consolidate all regularization terms, $\mathcal{L}{r}^{\mathcal{I}}$ and $\mathcal{L}_{r}^{\mathcal{U}}$, alongside the loss functions, $\mathcal{L}_{q}$ and $\mathcal{L}_{bpr}$, as defined in Eqs.~\eqref{eq:loss0}, \eqref{eq:loss2}, \eqref{eq:loss1}, and \eqref{eq:loss3}. These components are combined into the final objective function through a weighted aggregation process:
\begin{equation}
\mathcal{L} = \mathcal{L}_{bpr} + \alpha \mathcal{L}_{q} + \beta (\mathcal{L}_{r}^{\mathcal{U}} + \mathcal{L}_{r}^{\mathcal{I}}),
\label{eq:loss}
\end{equation}
where $\alpha$ is a hyperparameter regulating the contribution of the SSI module, and $\beta$ adjusts the degree of redundancy reduction introduced by RR.

The final objective function, $\mathcal{L}$, serves as the optimization criterion for the overall KSI framework, ensuring effective integration of the various components. 

\section{Experimental Settings}
\label{sec:es}

This section provides information on the datasets used in our paper, the evaluation metrics adopted for performance assessment, the state-of-the-art methods chosen for comparison, and the key details of implementation.

\begin{table}[]
\caption{\textbf{The statistics of datasets.}}
\centering
\begin{tabular}{ccccccc}
\toprule
Dataset                 &    & \multicolumn{2}{c}{\textbf{Sports}}  &    & \multicolumn{2}{c}{\textbf{Clothing}}        \\ \hline 
Modality  &    & Visual & Textual &  & Visual & Textual   \\   
Embed Dim &    & 4096 & 1024  & & 4096 & 1024     \\ \cline{1-1} \cline{3-4} \cline{6-7}  
User   &  & \multicolumn{2}{c}{35k}  &    & \multicolumn{2}{c}{39k}  \\ 
Item  &  & \multicolumn{2}{c}{18k}    &    & \multicolumn{2}{c}{23k}  \\  
Interactions   &  & \multicolumn{2}{c}{256k}  &    & \multicolumn{2}{c}{237k} \\ \bottomrule 
\end{tabular}
\label{tb:dataset}
\end{table}

\begin{table*}[t]
\caption{\textbf{Overall performance of KSI and our baselines on Sports and Clothing Datasets. $\dagger$ Statistically significant improvements (p-value $<$ 0.01).}}
\label{tab:overall_performance}
\centering
\small
\resizebox{0.98\textwidth}{!}{
\begin{tabular}{ccccccclcccccc}
\toprule
\multirow{2}{*}{Method} & 
\multicolumn{4}{c}{Sports} & & 
\multicolumn{4}{c}{Clothing} \\
\cmidrule(lr){2-5} \cmidrule(lr){7-10}
& Recall@10 & Recall@20 & NDCG@10 & NDCG@20 & & Recall@10 & Recall@20 & NDCG@10 & NDCG@20 \\
\midrule
MF-BPR & 0.0432 & 0.0653 & 0.0241 & 0.0298 & & 0.0187 & 0.0279 & 0.0103 & 0.0126 \\
LightGCN & 0.0569 & 0.0864 & 0.0311 & 0.0387 & & 0.0340 & 0.0526 & 0.0188 & 0.0236 \\
\hline
VBPR & 0.0558 & 0.0856 & 0.0307 & 0.0384 & & 0.0281 & 0.0415 & 0.0158 & 0.0192 \\
MMGCN & 0.0370 & 0.0606 & 0.0193 & 0.0254 & & 0.0218 & 0.0345 & 0.0110 & 0.0142 \\
GRCN & 0.0598 & 0.0915 & 0.0332 & 0.0414 & & 0.0424 & 0.0662 & 0.0223 & 0.0283 \\
LATTICE & 0.0620 & 0.0953 & 0.0335 & 0.0421 & & 0.0492 & 0.0733 & 0.0268 & 0.0330 \\
MMGCL & 0.0660 & 0.0994 & 0.0362 & 0.0448 & & 0.0438 & 0.0669 & 0.0239 & 0.0297 \\
SLMRec & 0.0663 & 0.0990 & 0.0365 & 0.0450 & & 0.0452 & 0.0675 & 0.0247 & 0.0303 \\ 
BM3 & 0.0656 & 0.0980 & 0.0355 & 0.0438 & & 0.0422 & 0.0621 & 0.0231 & 0.0281 \\
DiffMM & \underline{0.0670} & \underline{0.1014} & \underline{0.0376}& \underline{0.0458} & & \underline{0.0521} & \underline{0.0788} & \underline{0.0286} & \underline{0.0352} \\
\midrule
KSI & \textbf{0.0701}$\dagger$ & \textbf{0.1035}$\dagger$ & \textbf{0.0396}$\dagger$ & \textbf{0.0483}$\dagger$ & & \textbf{0.0547}$\dagger$ & \textbf{0.0807}$\dagger$ & \textbf{0.0300}$\dagger$ & \textbf{0.0366}$\dagger$ \\
Improv.  & 4.63\% & 2.07\%  & 5.32\% & 5.46\% & & 4.99\% & 2.41\% & 4.90\% & 3.98\%  \\ 
\bottomrule
\end{tabular}
}
\begin{flushleft}
\end{flushleft}
\end{table*}

\begin{table*}[t]
\centering
\caption{\textbf{Performance comparison of variants on different datasets.
The best result is in bold, and the gray background is the optimal result group. }}
\vspace{-5pt}
\setlength{\tabcolsep}{1.55mm}
\small
\begin{tabular}{ccccccclcccccc}
\toprule
\multirow{2}{*}{Variant} & \multicolumn{4}{c}{Sports}  &           & \multicolumn{4}{c}{Clothing}                    \\ \cline{2-5}  \cline{7-10}
& {Recall@10} & {NDCG@10} & {Recall@20}& {NDCG@20} & & {Recall@10} & {NDCG@10} & {Recall@20} & {NDCG@20} \\
\midrule
NGCF & 0.0374 & 0.0209 & 0.0695 & 0.0318 & & 0.0167 & 0.0089 & 0.0392 & 0.0169  \\
NGCF +\ SEI & 0.0581 & 0.0322 & 0.0867 & 0.0399 & & 0.0410 & 0.0227 & 0.0612 & 0.0279 \\
NGCF +\ SSI & 0.0558 & 0.0308 & 0.0838 & 0.0382 & & 0.0354& 0.0190 & 0.0557 & 0.0242 \\
\rowcolor{lightgray} NGCF +\ both  & 0.0586 & 0.0331 & 0.0879 & 0.0398 & & 0.0447 & 0.0242 & 0.0662 & 0.0297\\
\midrule
LightGCN & 0.0538 & 0.0308 & 0.0782 & 0.0369 & & 0.0314 & 0.0173 & 0.0470 & 0.0211  \\
LightGCN +\ SEI & 0.0679 & 0.0384 & 0.1000 & 0.0469 & & 0.0498& 0.0279 & 0.0736 & 0.0340 \\
LightGCN +\ SSI & 0.0645 & 0.0362 & 0.0948 & 0.0442 & & 0.0504 & 0.0279 & 0.0758 & 0.0344 \\
\rowcolor{lightgray} LightGCN +\ both  & \textbf{0.0701} & \textbf{0.0396} & \textbf{0.1035} & \textbf{0.0483} & & \textbf{0.0547} & \textbf{0.0300} & \textbf{0.0807} & \textbf{0.0366} \\

\bottomrule
\end{tabular}
\label{tab:results_1}
\end{table*}

\subsection{Datasets and Metrics}
\subsubsection{Datasets} 
We utilize two real-world public datasets used for multimodal recommendation: \textbf{Sports}, and \textbf{Clothing}~\footnote{Datasets are available at \url{https://jmcauley.ucsd.edu/data/amazon}} following the established protocols from LATTICE~\cite{LATTICE_zhang2021mining}. These datasets exemplify distinct recommendation scenarios with complementary multimodal characteristics. The 5-core filtering ensures data density while maintaining natural interaction sparsity. Our preprocessing pipeline aligns with MMGCN~\cite{MMGCN_wei2019mmgcn} for fair comparison.
Moreover, textual feature embeddings are produced using Sentence-BERT~\cite{datasets_reimers2019sentence} based on extracted information from product titles, descriptions, brands, and categorical data. Visual feature embeddings, with a dimensionality of $4096$, are generated from product images. Table~\ref{tb:dataset} provides detailed statistics, including embedding dimensionality.

Visual features were extracted from Amazon-provoded~\cite{ni2019justifying} binary image representations using ResNet-50 pretrained on ImageNet, generating 4096-dimensional embeddings through global average pooling of the final convolutional layer. For items with missing visual data, we employed mean imputation across the feature dimension. Textual features were generated through Sentence-BERT embeddings of concatenated product metadata including titles, descriptions, brands, and categorical information. Our feature processing follows established conventions in the field. Though many other ways could theoretically be utilized for visual features, we adopt the mainstream approach to ensure direct comparability with existing works.

\subsubsection{Evaluation Metrics} 
Three commonly used metrics are employed to evaluate the performance of top-$K$ recommendation results: Recall@$K$ (R@K), and Normalized Discounted Cumulative Gain@$K$ (N@K). The NDCG@$K$ is defined as:
\begin{equation}
NDCG@K = \frac{DCG@K}{IDCG@K},
\end{equation}
where $IDCG@K$ represents the ideal discounted cumulative gain, and
\begin{equation}
DCG@K = \sum_{i=1}^K \frac{rel_i}{\log(i+1)},
\end{equation}
with $rel_i$ denoting the relevance score of the $i$-th item.

Following the approach outlined in~\cite{BM3_zhou2023bootstrap}, we adopt an all-rank item evaluation strategy to assess accuracy. The reported results are averaged across all users in the test set to ensure robust evaluation.

\subsection{Baseline Methods} 
We compare KSI with several state-of-the-art (SOTA) recommendation approaches, categorized into two groups: collaborative filtering (CF) methods (\emph{i.e.}, MF, NGCF, and LightGCN) and deep content-aware recommendation models.

\textbf{Collaborative Filtering Methods:}
MF-BPR~\cite{MF_rendle2012bpr}: Optimizes Matrix Factorization using Bayesian Personalized Ranking (BPR) loss to capture implicit user-item interactions.
VBPR~\cite{VBPR_he2016vbpr}: Combines visual features and item embeddings in a Matrix Factorization framework; in this study, multimodal features replace the original visual-only inputs.
LightGCN~\cite{LightGCN_he2020lightgcn}: Employs a simplified graph-based model with lightweight graph convolution layers and layer aggregation.

\textbf{Deep Content-Aware Methods:} 
MMGCN~\cite{MMGCN_wei2019mmgcn}: Models modality-specific graphs and refines multimodal features through graph convolution, integrating them to construct user/item representations.
GRCN~\cite{GRCN_yinwei2021grcn}: Identifies false-positive feedback and prunes noisy edges in the interaction graph, thereby updating user/item representations.
LATTICE~\cite{LATTICE_zhang2021mining}: Learns modality-specific item-item structures, aggregates them into latent item graphs, and refines item representations via graph convolutions.
MMGCL~\cite{MMGCL_yi2022multi}: Integrates graph contrastive learning techniques using modality edge dropout and masking to enhance the recommendation model.
SLMRec~\cite{SLMRec_tao2022self}: Implements data augmentation strategies for multimodal content, such as noise perturbation and multimodal pattern augmentation. 
\textbf{BM3~\cite{BM3_zhou2023bootstrap}}: Eliminates negative sampling in BPR loss through bootstrapped contrastive views, reducing computational overhead while maintaining performance.
\textbf{DiffMM~\cite{DiffMM_jiang2024diffmm}}: Integrates graph diffusion models with cross-modal contrastive learning, dynamically aligning multimodal features with collaborative signals.
Our results demonstrate KSI's superiority across all categories, underscoring its ability to balance structural and semantic knowledge.

We implement variants based on MF, NGCF, and LightGCN, with LightGCN delivering the highest performance. All baseline models are reproduced under identical configurations and hardware settings as KSI. The reported results represent the best performance, chosen as the superior outcome between the original paper's findings and our reproduced results. 

\subsection{Implementation Details}
To ensure a fair comparison, we follow standard practices by utilizing the commonly pre-processed datasets provided in prior research~\cite{LATTICE_zhang2021mining}. The implementation of KSI is conducted using PyTorch~\cite{pytorch_paszke2017automatic} on NVIDIA RTX 3090 GPU. Consistent with the settings in LATTICE, the embedding dimension is fixed at 64, the Adam optimizer~\cite{adam_kingma2014adam} is employed with a learning rate of 0.0005, and the batch size is set to 1024. The model is trained for 50 epochs with parameters initialized using the Xavier initializer~\cite{Xavier_glorot2010understanding}. Additionally, the number of neighbors in $k$-nearest neighbors (as discussed in Section~\ref{cig}) is set to 10.

Hyperparameter tuning is conducted using the validation set. The number of layers is selected from 1, 2, 3, and 4, with 1 determined to be optimal. For the temperature parameter $\tau$ in the InfoNCE loss (Eq.\eqref{eq:loss1}), values are tested from 0.01, 0.05, 0.1, 0.5, and 1, with 0.1 yielding the best performance. The coefficient $\alpha$ in Eq.\eqref{eq:loss} is tuned across 0.01, 0.05, 0.1, 0.5, and 1, with 0.05 chosen for the Sports dataset and 0.1 for the Clothing dataset. Finally, $\beta$ in Eq.~\eqref{eq:loss} is evaluated over 0.1, 0.5, 1, 5, and 10, and a value of 1 is selected. 
\section{Experimental Results}
\label{sec:er}
In this section, we conduct extensive experiments to address the following Research Questions (RQs): 

\begin{itemize}[leftmargin=*] 
\item \textbf{RQ1:} What is the overall performance of KSI compared to SOTA methods? 
\item \textbf{RQ2:} What is the effectiveness of our proposed modules (SEI module and SSI module)? 
\item \textbf{RQ3:} How does RGNN perform, and should we enhance user embeddings? 
\item \textbf{RQ4:} Are our models sensitive to various hyperparameters? \end{itemize} 

\subsection{RQ1: Overall Results} 
We present the performance comparisons between KSI and various state-of-the-art (SOTA) CF and multimodal recommendation methods in Table~\ref{tab:overall_performance}. KSI integrates the CF framework LightGCN~\cite{LightGCN_he2020lightgcn} with the SEI and SSI modules. The results demonstrate that KSI consistently outperforms all baselines, achieving performance improvements ranging from 6.03\% to 14.31\% over the second-best models. Several critical insights can be drawn:

\noindent\textbf{(a)} Compared to CF-based approaches, content-aware methods exhibit superior performance overall, highlighting the potential of multimodal content to enhance recommendation accuracy.

\noindent\textbf{(b)} KSI surpasses all baselines across datasets by exploiting item-item structures instead of directly embedding multimodal content features, thereby mitigating the curse of knowledge. This is in contrast to methods like MMGCN and RGCN, which directly incorporate multimodal features into item representations.

\noindent\textbf{(c)} While LATTICE performs competitively, KSI achieves superior results by integrating both structural and semantic information from multimodal content, addressing the curse of knowledge through the SEI and SSI modules. Additionally, KSI's graph learning mechanism further enhances the representation's expressiveness.

\noindent\textbf{(d)} The performance improvement of KSI on the Baby dataset is relatively smaller. This may be attributed to the design of the core components (\ref{ref:rr}, \ref{ref:ssi}), which focus on minimizing redundancy and enabling soft integration of auxiliary knowledge. The Baby dataset, however, features minimal redundancy and limited auxiliary knowledge, reducing the scope for improvement.

\begin{figure*}[t!]
\centering
\subfigure[Study on RGNN.]{
\begin{minipage}[t]{0.22\linewidth}
\centering
\includegraphics[width=1.6in]{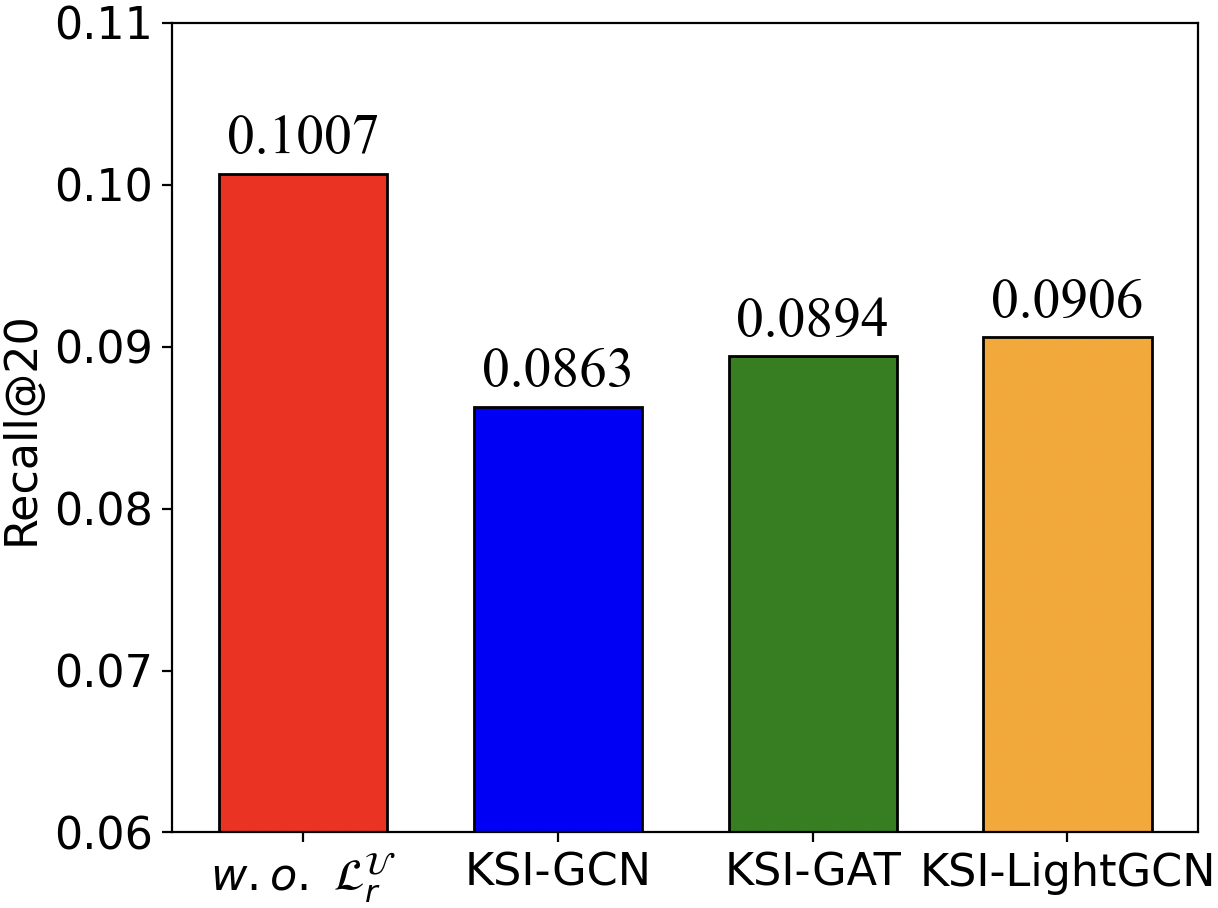}
\label{fig:a}
\end{minipage}%
}%
\subfigure[Study on $d$.]{
\begin{minipage}[t]{0.22\linewidth}
\centering
\includegraphics[width=1.6in]{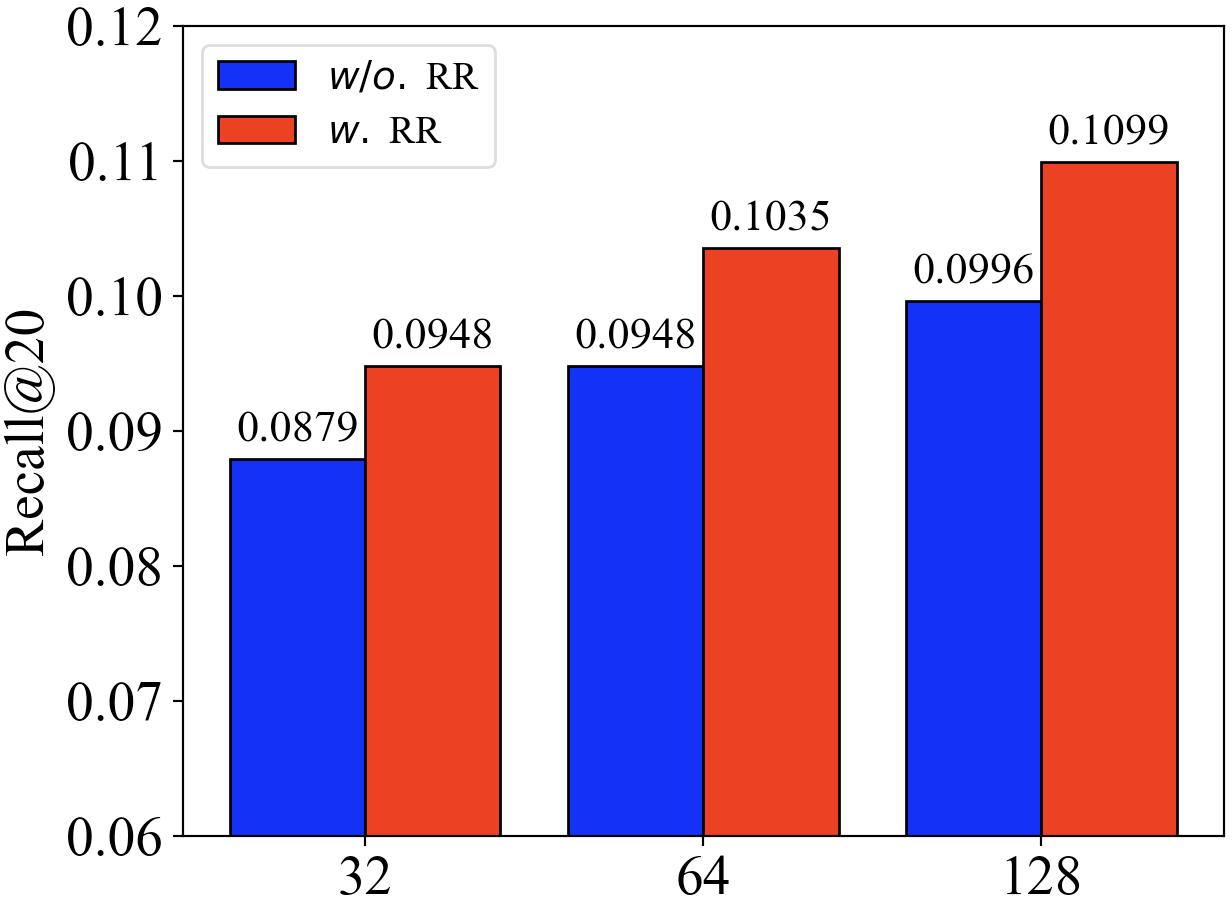}
\label{fig:b}
\end{minipage}%
}%
\subfigure[Study on $L$.]{
\begin{minipage}[t]{0.22\linewidth}
\centering
\includegraphics[width=1.6in]{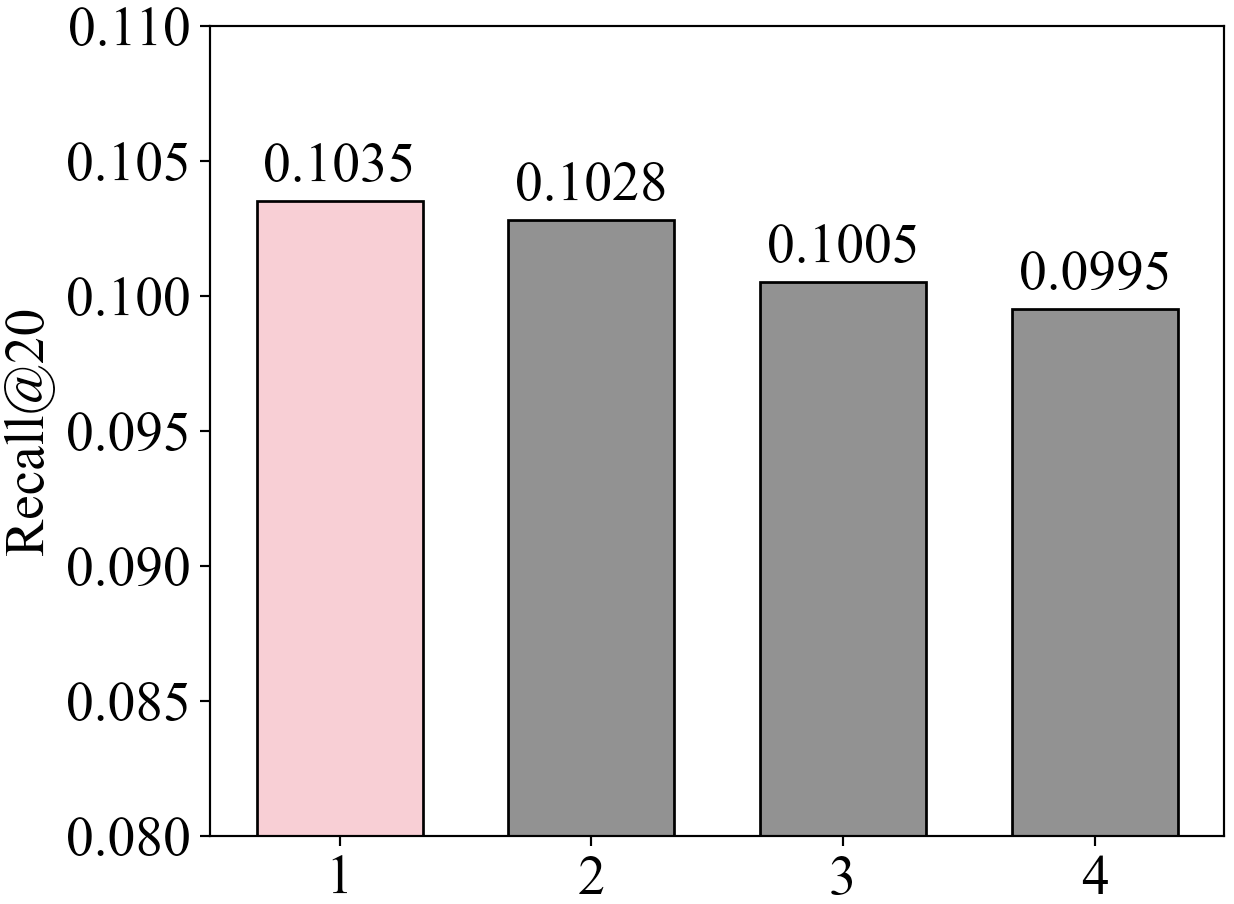}
\label{fig:c}
\end{minipage}
}
\subfigure[{Study on $K$.}]{
\begin{minipage}[t]{0.22\linewidth}
\centering
\includegraphics[width=1.8in]{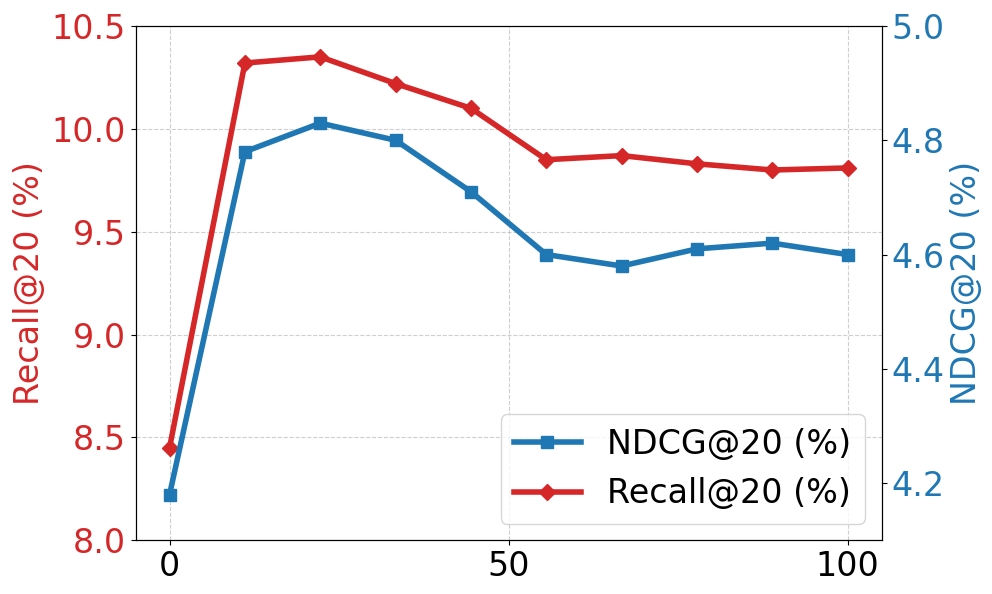}
\label{fig:e}
\end{minipage}
}
\caption{Performance comparison of our variants study in terms of Recall@20 on the Sports dataset.}
\end{figure*}

\begin{figure}
    \centering
    \includegraphics[width=2.5in]{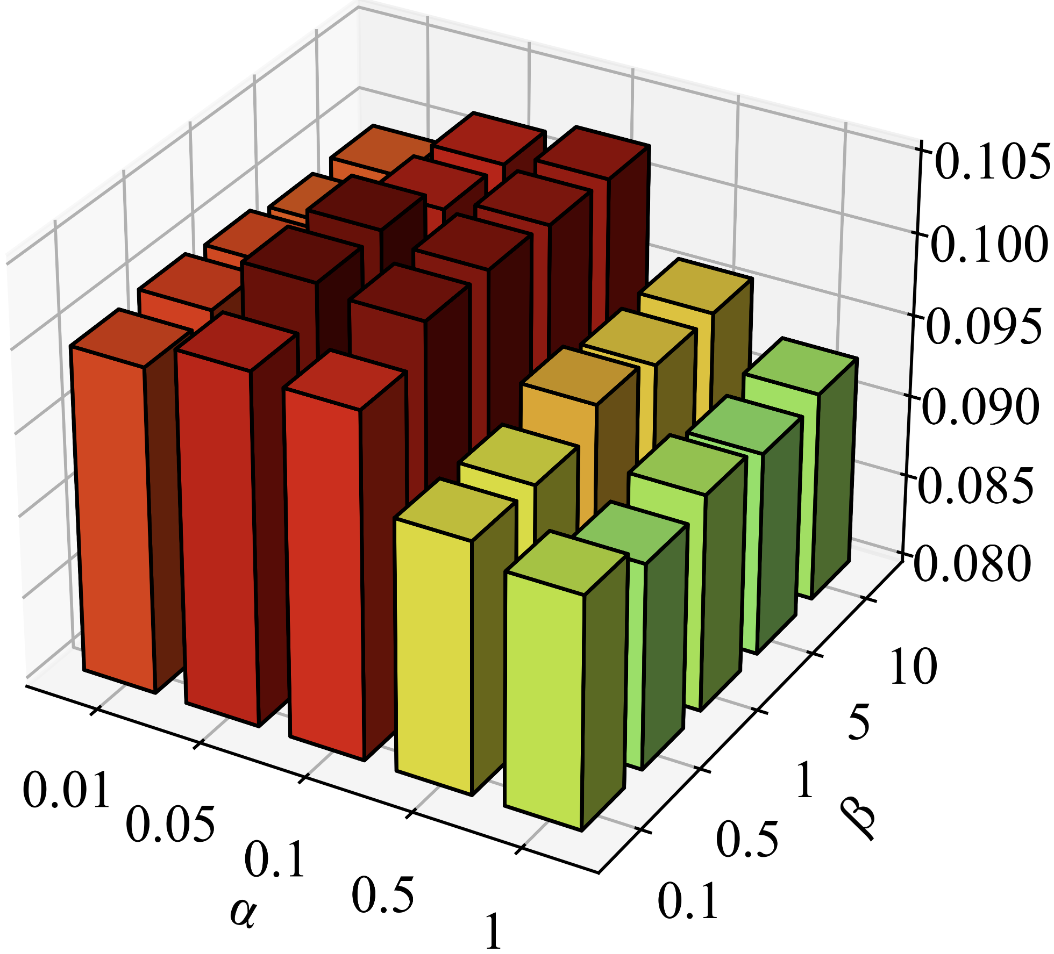}
    \caption{Study on $\alpha$ and $\beta$.}
    \label{fig:d}
\end{figure}
\subsection{RQ2: Effectiveness of Proposed Modules (SEI module and SSI module)}
To assess the portability and effectiveness of the SEI and SSI modules, we progressively integrate them into CF models and evaluate the resulting performance. Since the enhancements observed on the Clothing and Sports datasets are significantly greater than those on the Baby dataset, additional experiments focus on these two datasets. The experimental results, presented in Table~\ref{tab:results_1}, lead to the following observations:

\textbf{(a)} CF models augmented with the SEI module consistently demonstrate superior performance compared to their original versions, confirming the SEI module's portability and effectiveness. Remarkably, LightGCN integrated with SEI achieves performance levels that significantly exceed those of LATTICE, which also incorporates LightGCN, underscoring the advantages of RGNN and its capability for effective structural information transfer. 

\textbf{(b)} Incorporating the SSI module into CF models also results in performance gains over the original models, confirming the module's portability and utility. Furthermore, models augmented with the SSI module typically outperform state-of-the-art methods such as MMGCN and GRCN. In particular, LightGCN with SSI demonstrates superior performance compared to LATTICE, emphasizing the robustness of our feature retrieval task and the critical role of semantic information integration.

\textbf{(c)}  The highest performance is achieved when both SEI and SSI modules are integrated into baseline CF models. This underscores the complementary nature of structural and semantic information in multimodal recommendation systems. Experimental results demonstrate that KSI maintains superior performance with 2.07\%-5.46\% improvements across metrics, confirming its effectiveness against the latest advancements in multimodal recommendation.

\subsection{RQ3: Study of RGNN} To examine the performance of RGNN in comparison to other popular graph neural networks, we conduct comprehensive ablation experiments, as shown in Figure~\ref{fig:a}. Here, $\emph{w/o.}\ \mathcal{L}{r}^{\mathcal{U}}$ denotes the KSI variant without the $\mathcal{L}{r}^{\mathcal{U}}$ term. KSI-GCN, KSI-LightGCN, and KSI-GAT indicate variants where the RGNN layer in $\emph{w/o.}\ \mathcal{L}_{r}^{\mathcal{U}}$ is replaced with GCN, LightGCN, and GAT, respectively.

The results indicate that the variant with the RGNN layer significantly outperforms the others, underscoring RGNN’s superiority and showing that reducing redundancy to enhance expressiveness is beneficial for model performance. Furthermore, omitting the $\mathcal{L}_{r}^{\mathcal{U}}$ term results in performance degradation, highlighting the necessity of enhancing user embeddings. 

\subsection{RQ4: Hyperparameters} To assess the sensitivity of the model to variations in key hyperparameters, we perform ablation experiments on $d$, $L$, $\alpha$, $\beta$, $k$ and $K$. The results are presented in Figures~\ref{fig:b}, \ref{fig:c}, and \ref{fig:d}. Key observations include: 

\textbf{(a)} We examine KSI’s performance with varying embedding dimensions (32, 64, and 128) to analyze the RR process across different values of $d$. Results in Figure~\ref{fig:b} reveal that: 
\textbf{(i)}Performance improves significantly as the embedding dimension increases, indicating the challenge posed by rich multimodal information to representation expressiveness and underscoring the importance of reducing redundancy. 
\textbf{(ii)} In all dimension settings, the RR process consistently enhances performance, demonstrating its effectiveness in reducing redundancy. Notably, this process enables better performance even at smaller embedding sizes.

\textbf{(b)} To evaluate the impact of the number of RGNN layers, we vary $L$ over the range $[0, 1, 2, 3, 4]$. As shown in Figure~\ref{fig:c}, KSI achieves its lowest performance at $L = 0$, highlighting the importance of structure information. The best results are observed when $L$ is set to 1.

\textbf{(c)} Both $\alpha$ and $\beta$ are essential to KSI’s performance. We tune $\alpha$ over $[0.01, 0.05, 0.1, 0.5, 1]$ and $\beta$ over $[0.1, 0.5, 1, 5, 10]$. Figure~\ref{fig:d} shows that KSI achieves optimal performance on the Sports dataset when $\alpha$ is set to 0.05 and $\beta$ to 0.5. 

\textbf{(d) Figure~\ref{fig:e} shows that KSI achieves stable performance across varying \( k \)-NN sparsification values, with optimal results at \( k=10 \).}

\section{Conclusion}
We tackle the pervasive \textit{curse of knowledge} challenge in existing multimodal recommendation approaches by introducing KSI, a framework designed to balance the effective utilization of multimodal features while mitigating this issue. KSI incorporates the SEI and SSI modules, which work synergistically to enhance recommendation performance. A key component of the framework is the RGNN, embedded within the SEI module, which is specifically designed to mine item-item structural relationships and improve representation expressiveness. Comprehensive experimental evaluations demonstrate that KSI consistently surpasses state-of-the-art methods, validating the portability and efficacy of the SEI and SSI modules.

Our future work will focus on extending the application of RGNN to broader contexts and investigating additional strategies for knowledge soft integration to further enhance the robustness of multimodal recommendation systems.

\section*{Acknowledgments}
This work was supported by National Natural Science Foundation of China~(Grant No. 92467204 and 62472249), and Shenzhen Science and Technology Program (Grant No. JCYJ20220818101014030 and KJZD20240903102300001).
\bibliographystyle{IEEEtran}
\bibliography{IEEEabrv,sample-base}

\end{document}